\documentclass[prd,twocolumn,nofootinbib,notitlepage,10pt,aps]{revtex4-2}
\pdfoutput=1
\usepackage[utf8]{inputenc}

\usepackage{graphicx}
\usepackage{amsmath,amsfonts,amssymb}
\usepackage{color}

\usepackage{verbatim}
\usepackage{enumitem}

\usepackage{orcidlink}
\usepackage{booktabs}

\usepackage[normalem]{ulem}
\usepackage{multirow}
\usepackage{arydshln}
\usepackage{nicematrix,tikz}

\usepackage{ulem}
\usepackage{pdfsync}
\usepackage{epsfig}
\usepackage{epstopdf}

\usepackage{subfigure}
\usepackage{comment}

\usepackage{slashed}
\usepackage{physics}

\def\m{\mu}

\def\s{\sigma}

\usepackage{braket}
\usepackage{colortbl}

\newcommand{\be}{\begin{equation}}
\newcommand{\ee}{\end{equation}}
\newcommand{\bea}{\begin{equation} \begin{aligned}}
\newcommand{\eea}{\end{aligned} \end{equation}}

\newcommand{\Mpl}{M_{P}}

\newcommand{\td}{{\rm d}}

\usepackage{hyperref}
\hypersetup{
    colorlinks,
    linkcolor={violet},
    citecolor={teal},
    urlcolor={teal}
}

\begin{document}

\title{Quintessence and phantoms in light of DESI 2025}

\author{Ioannis D. Gialamas\orcidlink{0000-0002-2957-5276}}
\email{ioannis.gialamas@kbfi.ee}
\author{Gert H\"utsi\orcidlink{0000-0002-9322-004X}}
\email{gert.hutsi@kbfi.ee}
\author{Martti Raidal\orcidlink{0000-0001-7040-9491}}
\email{martti.raidal@cern.ch}
\author{Juan Urrutia\orcidlink{0000-0002-6035-6610}}
\email{juan.urrutia@kbfi.ee}
\author{Martin Vasar\orcidlink{0009-0003-3514-1575}}
\email{martin.vasar@ut.ee}
\author{Hardi Veerm\"ae\orcidlink{0000-0003-1845-1355}}
\email{hardi.veermae@cern.ch}

\affiliation{Keemilise ja Bioloogilise F\"u\"usika Instituut, R\"avala 10, 10143 Tallinn, Estonia}

\begin{abstract}
We analyze DESI BAO, CMB, and supernova data to explore the physical origin of the DESI indication for dynamical dark energy. Beyond the standard linear parametrization, $w(a) = w_0 + (1-a)w_a$, we explore truncated alternatives and quintessence models. We conclude that there is compelling evidence for dark energy to be decaying in the late universe, but the evidence for a phantom behavior, i.e., $w<-1$ at some redshift, is less significant. Models without phantom behavior are compatible with the data at the $2\sigma$ confidence level. Furthermore, we examine a concrete, theoretically well-motivated quintessence scenario with a Higgs-like potential, allowing for a direct comparison with parametrized approaches and testing its consistency with current observations. This framework enables a broader investigation of late-time cosmic evolution and reveals a $93.8\%$ preference for a future transition into an anti-de Sitter space, which may ultimately lead to a cosmological collapse of our Universe.
\end{abstract}

\maketitle

\section{Introduction}

Nearly a year after its first data release (DR1)~\cite{DESI:2024mwx}, the Dark Energy Spectroscopic Instrument (DESI) has published its second data release (DR2)~\cite{DESI:2025zgx}, based on three years of observations. This release includes baryon acoustic oscillation (BAO) measurements from millions of galaxies and quasars, providing precise constraints on the transverse comoving distance and the Hubble radius across seven redshift bins spanning the range $0.1<z<4.2$. One of the most significant results from the DESI BAO observations is the emerging tension with the $\Lambda$CDM model, which assumes that the dark energy (DE) is described by a cosmological constant and therefore has an equation of state fixed at $w = -1$ for all redshifts. While the DR1 results suggested that this discrepancy might be driven by the combination of BAO data with supernova (SN) distance moduli measurements, the new data indicate that even when the SN data are excluded, the combination of DESI BAO measurements with cosmic microwave background (CMB) data from the Planck satellite~\cite{Planck:2018vyg} shows a preference for dynamical DE. More specifically, when adopting the standard Chevallier-Polarski-Linder (CPL) parametrization~\cite{Chevallier:2000qy, Linder:2002et} of the DE equation of state $w(a) = w_0+w_a (1-a)$, this model is preferred over $\Lambda$CDM at $3.1\s$ for the combination of DESI BAO and CMB data, and at $4.2\s$ when SN data from Dark Energy Survey Year 5 (DESY5)~\cite{DES:2024jxu} are also included. The $\Lambda$CDM model appears to be in serious tension with the data, and explanations for this tension have been offered in light of both the DR1~\cite{Wang:2024dka,Allali:2024cji,Muursepp:2024mbb,Yin:2024hba,Wang:2024hks,Colgain:2024xqj,Wang:2024rjd,Carloni:2024zpl,Wang:2024pui,Luongo:2024fww,Mukherjee:2024ryz,Dinda:2024kjf,Wang:2024hwd,Heckman:2024apk,Gialamas:2024lyw,Notari:2024rti,Giare:2024gpk,Dinda:2024ktd,Jiang:2024xnu,Efstathiou:2024xcq,Perivolaropoulos:2024yxv,RoyChoudhury:2024wri,Giare:2024ocw,Muralidharan:2024hsc,Chan-GyungPark:2024brx,Notari:2024zmi,Colgain:2024mtg,Gao:2024ily,Liu:2024yib,Lewis:2024cqj,Luongo:2024zhc,Alfano:2024fzv,Chan-GyungPark:2025cri,Huang:2025som,Giare:2025pzu,Mukherjee:2025fkf,Shah:2024gfu,Li:2024bwr,Bansal:2025ipo} and DR2~\cite{Ormondroyd:2025iaf,Moghtaderi:2025cns,Anchordoqui:2025fgz,Paliathanasis:2025dcr,Moffat:2025sik,Ye:2025ulq,Chaussidon:2025npr,Kessler:2025kju,Colgain:2025nzf,Cortes:2025joz,Scherer:2025esj,Specogna:2025guo,Cheng:2025lod,Efstathiou:2025tie,Ling:2025lmw,Gonzalez-Fuentes:2025lei} DESI results\footnote{Another notable tension is the mismatch between the sum of neutrino masses inferred from oscillation experiments and those favored by the combined CMB, SN, and DESI BAO data~\cite{Craig:2024tky,Wang:2024hen,Jiang:2024viw,Reboucas:2024smm,DESI:2025ejh,Capozzi:2025wyn,RoyChoudhury:2025dhe}.}. However, modified recombination scenarios, in which the evolution of the ionization fraction differs from the standard case, have been shown to not require dynamical DE and also predicts a higher value for the Hubble constant~\cite{Lynch:2024hzh,Mirpoorian:2025rfp}. Additionally, allowing for a higher optical depth to reionisation -- the least well-constrained CMB parameter -- alleviates the need for dynamical DE~\cite{Sailer:2025lxj}.

Models that introduce ultralight axionlike fields as dynamical DE -- some of which also attempt to account for the observed isotropic cosmic birefringence~\cite{Minami:2020odp,Komatsu:2022nvu} -- have been investigated in~\cite{Luu:2025fgw,Lee:2025yvn,Nakagawa:2025ejs,Urena-Lopez:2025rad,Lin:2025gne}. Rather than focusing on evolving DE, alternative approaches have been considered that aim to explain the new observational results with models featuring nonstandard dark matter evolution~\cite{Kumar:2025etf,Wang:2025zri,Chen:2025wwn,Abedin:2025dis}.
Possible interactions between DE and dark matter, in light of DESI observations, have been thoroughly discussed~\cite{Giare:2024smz,Li:2024qso,Li:2024hrv,Chakraborty:2025syu,Khoury:2025txd,Shah:2025ayl,Silva:2025hxw,You:2025uon,Pan:2025qwy}.

Another key aspect of the latest results is the indication of a phantom crossing scenario, i.e., $w(z)<-1$ at sufficiently high redshifts. This scenario is unavoidable when fitting the data with the CPL parametrization~\cite{DESI:2025zgx}. Moreover, although in the CPL parametrization, the low $z$ observations can drive the preference for a phantom solution at higher $z$~\cite{Gialamas:2024lyw}, the DESI Collaboration claims the current data strongly prefer models featuring a phantom crossing, regardless of the parametrization~\cite{DESI:2025fii}. While this behavior poses theoretical challenges for most straightforward scalar-field models of quintessence~\cite{Tada:2024znt,Berghaus:2024kra,Shlivko:2024llw,Bhattacharya:2024hep,Ramadan:2024kmn,Gialamas:2024lyw,Wolf:2024eph,Bhattacharya:2024kxp,Payeur:2024dnq,Berbig:2024aee,Taylor:2024whh,Aboubrahim:2024cyk,Park:2024ceu,Mukherjee:2025myk,Lu:2025gki,Borghetto:2025jrk,Shlivko:2025fgv,Akrami:2025zlb,Lin:2025gne,Dinda:2025iaq,deSouza:2025rhv,VanRaamsdonk:2025wvj,Bayat:2025xfr,Wang:2025dtk,Cline:2025sbt}, it can be accommodated within modified theories of gravity~\cite{Yang:2024kdo,Escamilla-Rivera:2024sae,Ye:2024ywg,Wolf:2024stt,Ye:2024zpk,Ferrari:2025egk,Yang:2025kgc,Wolf:2025jlc,Pan:2025psn,Yang:2025mws,Li:2025cxn,Wolf:2025jed,Paliathanasis:2025hjw,Cai:2025mas,Bahamonde:2025eov}, although the analysis of the DESI DR1 results indicated that general relativity remains the most likely theory of gravity consistent with observations~\cite{Ishak:2024jhs,Specogna:2024euz}.

Since the primary focus of this paper is on quintessence models, it is useful to first examine in more detail existing works on such models, as well as on modified gravity scenarios with nonminimal couplings that can give rise to phantom behavior. (a) \emph{quintessence models:}
in light of both DESI DR1 and DR2, several studies have examined scalar field models in comparison to the CPL parametrization. In Ref.~\cite{Shlivko:2024llw}, the exponential and hilltop models were mapped to CPL parameters. The exponential model was fitted in Refs.~\cite{Bhattacharya:2024hep,Ramadan:2024kmn}, but CPL provided a better fit. The hilltop model was tested in Refs.~\cite{Wolf:2024eph,Cline:2025sbt}, with CPL again performing better. Even the Higgs-like model with vanishing vacuum energy analyzed in Ref.~\cite{Bhattacharya:2024kxp} did not outperform CPL. Using DESI DR2 data, Ref.~\cite{Akrami:2025zlb} shows that the exponential model fits worse than CPL. Including spatial curvature improved the fit, but it remained less favourable than CPL. (b) \emph{phantom models:} in Ref.~\cite{Ye:2024ywg}, a nonparametric reconstruction approach of $w(z)$ indicated a crossing of the phantom divide. Alongside this, an exponential potential with nonminimal coupling was also considered, though it still resulted in a slightly poorer fit compared to CPL. A more elaborate model, incorporating both an exponential potential, a constant vacuum energy term, and a nonminimal coupling to gravity, was explored in~\cite{Wolf:2024stt}; while it yielded a better fit than CPL, this improvement came at the cost of additional free parameters. In contrast, Refs.~\cite{Wolf:2025jlc,Wolf:2025jed} demonstrated that a model featuring nonminimal coupling and a potential of the form $\sim V_0+\phi -\phi^2$ provided a better fit than CPL. In any case, it is worth noting that although the presence of nonminimal couplings can lead to phantom behavior, which seems to be favored by the data, it also modifies the value of the effective gravitational constant, with consequences that are well known in the literature~\cite{Will:2014kxa} (see also Ref.~\cite{Linder:2025zxb} for a recent discussion arguing that this type of interaction does not provide a satisfactory solution to the dynamical DE issues).

In this paper, we show that the interpretation of a phantom equation of state is not necessarily correct. In particular, phantom DE can be an artifact of the parametrization, if the data can be fitted well by simple models that do not invoke phantoms~\cite{Gialamas:2024lyw}. Specifically, we start our investigations with a three-parameter generalisation of the CPL -- the \textit{truncated CPL} -- that imposes a lower bound on the CPL equation of state, $w(z) \geq w(z\rightarrow\infty)\equiv w_{\infty}$. We find that the combined data are marginally consistent with $w_{\infty} \approx -1$, allowing a constant DE solution at high redshifts. A similar conclusion was recently reached in~\cite{Nesseris:2025lke,Akrami:2025zlb} for a different class of CPL extensions. 

Moreover, we will show that the two-parameter sub-model with $w_{\infty} = -1$ as well as the two-parameter \textit{quintessence} parametrization~\cite{Dutta:2008qn,Chiba:2009sj,Dutta:2009yb,Chiba:2009nh,Dutta:2009dr}, can fit the data well, without exhibiting any phantom behavior.
This motivates us to explore a well-motivated scalar field model that may support a nonphantom Universe. Specifically, we consider a Higgs-like potential, as observational data appear to favour scenarios with an effective negative mass-squared term. In this setup, the Universe is currently transitioning from the symmetric phase to a spontaneously broken one. This framework not only provides a consistent dynamical setting but also offers insight into the future evolution of the Universe, including the possibility of a cosmological collapse~\cite{Steinhardt:2001st,Felder:2002jk,Heard:2002dr,Kallosh:2002gf}.

The structure of the paper is as follows: In Sec.~\ref{sec:models}, we present the key features of the various equation of state parametrizations, along with the explicit Higgs-like quintessence model. In Sec.~\ref{sec:meth_res}, we present the numerical analysis of the observational data fits, with additional technical details provided in Appendix~\ref{appendixa}. Finally, our findings are summarized and conclusions drawn in Sec.~\ref{sec:concl}.

\section{Modeling dynamical dark energy and quintessence}
\label{sec:models}

In this section, we consider two complementary approaches to modeling dark energy: (i) parametrizations of the equation of state parameter 
$w(z)$, and (ii) an explicit quintessence model based on a minimally coupled scalar field. Parametrized forms of $w(z)$, such as the commonly used CPL parametrization and its extensions, are employed for their simplicity and flexibility, allowing us to explore the general behavior of DE and assess possible preferences for phantomlike behavior ($w<-1$). Additionally, we consider a quintessence-inspired parametrization based on the evolution of a scalar field. In the second part of the analysis, we move to an explicit model of quintessence with a Higgs-like potential, enabling a direct comparison with the parametrized approaches and an evaluation of its consistency with the latest observational data.

Assuming standard general relativity in a spatially flat Friedmann–Robertson–Walker (FRW) background, the first Friedmann equation takes the form
\be
\label{eq:hubble}
    \frac{H^2}{H_0^2} = \Omega(z) +\Omega_{\rm DE}\frac{\rho_{\rm DE} (z)}{\rho_{{\rm DE},0}}\,.
\end{equation}
The function $\Omega(z) = \Omega_m (1+z)^3 + \Omega_{\gamma} (1+z)^4 + \Omega_{\nu}(z)$ represents the total normalized energy density of matter, radiation and neutrinos at redshift $z$ and $\Omega_i=\rho_{i,0}/(3\Mpl^2H_0^2)$ denotes the present-day density parameter for the corresponding component $i$, with $\rho_{i,0}$ being the present-day energy density. Note also that $\Omega_m=\Omega_b+\Omega_c$. The energy density of neutrinos is included following Ref.~\cite{WMAP:2010qai}. The time dependence of the energy density of the DE component $\rho_{\rm DE} (z)$ is the main subject of this study. As discussed above, we will consider two approaches here:
\begin{enumerate}[label=\roman*.,leftmargin=*]
    \item a generic parametrization of the equation of state $w(z)$ of dynamical DE.
    
    \item quintessence in which the dynamics of DE is determined by the shape of the potential.
    
\end{enumerate}
In the first case, given an equation of state parameter $w(z)$, the energy density of DE evolves as
\be
    \rho_{\rm DE} (z) =\rho_{{\rm DE},0}\, e^{3\int_0^z \frac{1+w(z')}{1+z'}{\rm d}z'}\,,
\ee
where $\rho_{{\rm DE},0}$ is its current value.  We will consider parametrizations of $w(z)$ that can approximately mimic quintessence. Although our focus is on models of minimally coupled thawing quintessence, which generally does not exhibit phantom behavior at high $z$, we will also consider parametrizations that permit $w(z) < -1$ in order to test to what degree phantom behavior is preferred by the data.

\subsection{Parametrizations of \texorpdfstring{$w$}{w}}
\label{sec:params}

\textit{Truncated CPL} -- One of the simplest and most commonly used ways to model dynamical DE is the CPL parametrization~\cite{Chevallier:2000qy,Linder:2002et}, which is linear in the scale factor
\be
    {\rm CPL}: \qquad w(a) = w_0 + (1-a) w_a\,.
\ee
It interpolates between $w_0$ at present ($a=1$) and $w_0+w_a$ in the asymptotic past ($a=0)$. However, one of its drawbacks is that the dynamics at low and high $z$ are inevitably connected: a sufficiently rapid decay of DE ($w_a < -w_0$) at the lowest redshifts will imply a phantom behavior ($w<-1$) of DE in the past.

Therefore, to decouple the low and high $z$ expansion, the CPL ansatz must be extended\footnote{Extensions of the CPL parametrization beyond the linear approximation by including higher-order terms are discussed in~\cite{Nesseris:2025lke,Akrami:2025zlb}.}. A simple one-parameter extension, which also accommodates thawing quintessence approximately, is the truncated CPL
\be
   \mbox{tCPL}: \quad  w(a) = \max(w_0 + (1-a) w_a, w_{\infty})\,.
\ee
When $w_0 + w_a < w_{\infty} < w_0$, then the DE will have a constant equation of state parameter $w_{\infty}$  when $z > z_t \equiv (w_{\infty}-w_0)/(w_0+w_a - w_{\infty})$. 
A more constrained variant is the tCPL(-1) model, defined as
\be
   \mbox{tCPL(-1)}: \quad  w(a) = \max(w_0 + (1-a) w_a, -1)\,, 
\ee
which is equivalent to the tCPL with $w_{\infty} = -1$ and roughly resembles thawing quintessence, models in which DE is described by a scalar field initially frozen with an equation of state close to $w = -1$ that slowly evolves away from this value as the Universe expands and the Hubble friction decreases.

In the following analysis, we parametrize $w(a)$  for both the tCPL and tCPL(-1) models, considering $\log_{10}z_t$ and $w_a$ as free parameters in separate cases. Table~\ref{table:Priors} in Appendix~\ref{appendixa} lists the uniform priors of the parameters adopted in the analysis.

\textit{Dutta-Scherrer-Chiba (DSC) parametrization} --- In scenarios of thawing quintessence, at high $z$, the quintessence field is frozen at some value $\phi_0$ away from the minimum of the potential and begins to roll towards the minimum at late times, effectively behaving as dynamical DE that transitions from $w = -1$ to $w > -1$ at the present~\cite{Tsujikawa:2013fta}. If the field excursion is small enough, then its potential can always be expanded as
\begin{eqnarray}
\label{eq:Vphi_expansion}
    V(\phi) 
   &=& V(\phi_0) 
    + V'(\phi_0) (\phi - \phi_0) 
    \nonumber\\
   &+& \frac{1}{2}V''(\phi_0) (\phi - \phi_0)^2 
    + \mathcal{O}(\phi - \phi_0)^3\,.
\end{eqnarray}
The evolution of the equation of state parameter can be approximated by\footnote{The DSC parametrization in the context of DESI observations has been considered in Refs.~\cite{Tada:2024znt,Gialamas:2024lyw,Bhattacharya:2024kxp,Shlivko:2025fgv}.}~\cite{Dutta:2008qn,Chiba:2009sj}
\be\label{eq:w_quint}
    w(z)\simeq-1+(1+w_0)(1+z)^{3(1-K)}\mathcal{F}(z),
\ee
with
\begin{widetext}
\be
 \mathcal{F}(z)=\left[ \frac{(K-F(z))(F(z)+1)^K+(K+F(z))(F(z)-1)^K}{(K-\Omega_{\phi}^{-1/2})(\Omega_{\phi}^{-1/2}+1)^K+(K+\Omega_{\phi}^{-1/2})(\Omega_{\phi}^{-1/2}-1)^K}
    \right]^2\,, \qquad F(z)=\sqrt{1+(\Omega_{\phi}^{-1}-1)(1+z)^{3}}\,.
\ee
\end{widetext}
Here, $\Omega_{\phi}$ denotes the present-day energy density parameter of the scalar field $\phi$. The above formulation assumes a spatially flat Universe, such that $\Omega_{\phi} + \Omega_m = 1$. Eq.~\eqref{eq:w_quint} was initially derived assuming a vanishing linear term in~\eqref{eq:Vphi_expansion}~\cite{Dutta:2008qn} and later shown to work also for the general second-order expansion~\cite{Chiba:2009sj}. The parameter $K$ is defined slightly differently in these scenarios: it is a function of the second derivative of the potential at the local maximum in the former case and the initial value $\phi_i$ in the latter case. To eliminate this ambiguity, we estimate the parameter $K$ using
\be\label{eq:K}
    K = \left.\sqrt{1 - \frac{4}{3} \frac{V''(\phi)}{V(\phi) - V'(\phi)^2/(2V''(\phi))}}\right|_{\phi=\phi_i}\,.
\ee
The additional $V'$ term in the denominator accounts for the fact that the initial field $\phi_i$ is not at the local maximum, and it corresponds exactly to the expression given in Eq.~\cite{Dutta:2008qn} now parametrized in terms of $\phi_i$.

Given the relatively unrestrictive nature of the underlying assumptions, this parametrization covers a broad class of quintessence models. The most constraining assumption is that $\rho_\phi$ is approximately constant, which suggests that the ansatz captures quintessence worse when $w_0$ deviates significantly from $-1$.
\setlength{\tabcolsep}{4.2pt}
\renewcommand{\arraystretch}{1.8}
\begin{table*}[t!]
  \centering
\small  
  \begin{tabular}{|c|ccccccccc|}
    \hline 
    \hline 
    \\[-0.67cm]   
    & Model & Prior &  $\Delta \chi^2$ & $\Delta$AIC & $\Omega_m$ & $w_0$ & $w_a$ or $K$ & $z_t$ & $w_{\infty}$
    \\
    \hline
    \\[-0.67cm] 
    \multirow{3}{*}{\rotatebox[origin=c]{90}{}} & \bf{$\Lambda$CDM} & &  $0.0$ & $ 0.0$ & $0.302$ & --- & --- & --- & --- \\
   & & & &  & {\scriptsize $[0.295, 0.309]$} & --- & --- &  ---& --- \\
    \cdashline{1-10}[5pt/3pt]
    \\[-0.65cm] 
    \multirow{5}{*}{\rotatebox[origin=c]{90}{\textbf{Phantomlike}}} & \bf{CPL} & &  $-17.2$ & $-13.2$ & $0.317$ & $-0.770$ & $-0.782$ & --- & --- 
    \\
   & & & &  & {\scriptsize$[0.306, 0.328]$} & {\scriptsize$[-0.881, -0.651]$} & {\scriptsize $[-1.30, -0.34]$ } & ---  & --- 
   \\
    & \bf{tCPL} & &  $-17.2$ & $-11.2$ & $0.317$ & $-0.770$ & $-0.782$ & $6.45$ & $-1.45$ 
    \\
    &   &  $w_a$ &  & & \scriptsize$[0.300, 0.340]$ & \scriptsize$[-0.866, 1.89]$ & \scriptsize$[-5100, -0.081]$ & \scriptsize$[0.00007, 17.9]^*$ & \scriptsize$[-1.55, -0.90]$ 
    \\
    & & $\log_{10}{z_t}$ &  & & \scriptsize$[0.305, 0.328]$ & \scriptsize$[-0.941, -0.164]$  & \scriptsize$[-9.52, -0.035]^*$ & \scriptsize$>0.103$ &
    \scriptsize$[-1.81, -0.97]$ 
    \\
    \cdashline{1-10}[5pt/3pt]
    \multirow{8}{*}{\rotatebox[origin=c]{90}{\textbf{Quintessencelike}}} & \bf{tCPL(-1)} & &  $-13.5$  & $-9.5$ & $0.323$ & $-0.064$ & $-11.3$ & $0.090$ & -1 
    \\
    & & $w_a$ &  &  & {\scriptsize$[0.305, 0.346]$} & {\scriptsize$-0.233\geq$} & {\scriptsize$[-138, -1.02]$} & {\scriptsize$ [0.0009, 0.0962]$*} & {\scriptsize-1} 
    \\  
    & & $\log_{10}{z_t}$ &  & & {\scriptsize$[0.307, 0.341]$} & {\scriptsize$[-0.90, 1.78]$} & {\scriptsize$[-73.0, -0.09]$*} & {\scriptsize$[0.001, 0.372]$} & {\scriptsize-1} \\
    & \bf{DSC} & &  $-13.3$  & $-9.3$ & $0.326$ & $0.445$ & $14.8$ & --- & --- 
    \\
    &  & & & & {\scriptsize$[64.0, 67.1]$} & {\scriptsize$-0.451 <$} & {\scriptsize$[4.7, 35.2]$} & ---  &  --- 
    \\  
    \cline{2-10}
    & &   &  &  &  & $V_0/(H_0^2\Mpl^2)$ & $-m^2/H_0^2$ & $\log_{10}(v_\phi/\Mpl)$ & $V(z=0)/(H_0^2\Mpl^2)$ 
    \\
    \cline{2-10}
    \\[-0.65cm]
   & \bf{Higgs-like} &  &   $-13.5$ & $-7.5$ & $0.320$ & $2.27$ & $2950$ & $-1.41$  & {$1.21^*$}
    \\
    & & & &  & {\scriptsize $[0.311, 0.374]$} & {\scriptsize$[2.19, 2.65]$} & {\scriptsize$251<$} & \scriptsize$-1.11 < $  & \scriptsize{$[-9.30, 2.04]^*$}
    \\
    \hline
    \hline
  \end{tabular} 
\caption{Results of the maximal likelihood analysis of the CMB+DESI+DESY5 datasets for the CPL, tCPL, tCPL(-1), DSC and Higgs-like models. For each model parameter, we show the best-fit value and the 95\%  credible intervals. For the tCPL and tCPL(-1) ansatz, we show results using uniform priors in both $\log_{10}{z_t}$ and $w_a$ for which the credible intervals differ. The choice of the varied parameter with a flat prior is indicated in the second column. The asterisks indicate derived quantities. }
  \label{table:CredibleIntervals}
\end{table*}

\subsection{Higgs-like quintessence} 

\begin{figure*}
    \centering
    \includegraphics[width=1\linewidth]{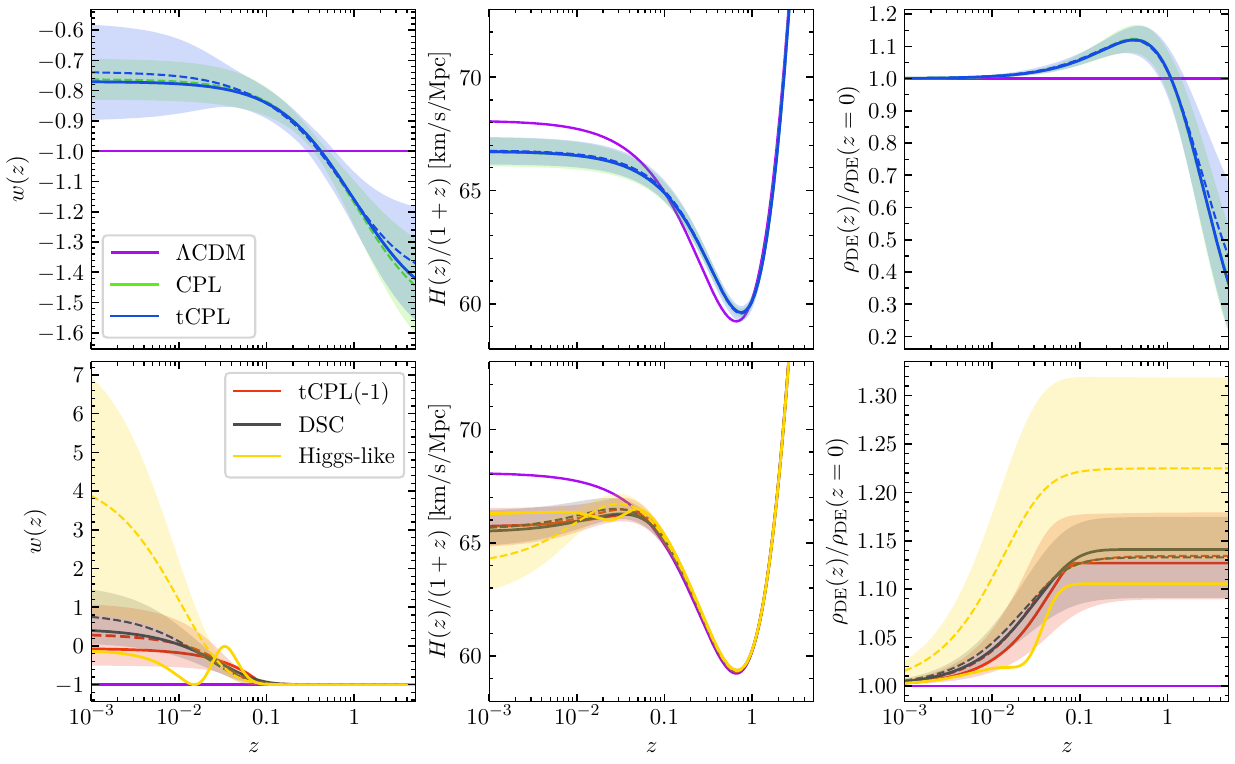}
    \caption{Evolution of $w(z)$, $H(z)/(1+z)$, and $\rho_{\rm DE}(z)$ for the best-fits (solid lines) and sample means (dashed lines), with the $1\sigma$ spread indicated by the shaded regions. The upper panels show models exhibiting phantom behavior at high-$z$, while the lower panels show quintessencelike scenarios of Table~\ref{table:CredibleIntervals}. The $\Lambda$CDM model is shown throughout as solid purple lines. }
    \label{fig:evol_z}
\end{figure*}

In this section, we move beyond the commonly used parametrizations of the dark energy equation of state typically found in the literature and instead adopt a more fundamental, physically motivated model based on quintessence. Specifically, we consider a canonically normalized scalar field $\phi$ minimally coupled to gravity through the Einstein-Hilbert action. The total action is given by 

\be
    \mathcal{S} 
    \!=\! \int{\rm d}^4x \sqrt{-g} \left(\frac{\Mpl^2}{2}R \!-\! \frac{(\partial_\m \phi)^2}{2} \!-\! V(\phi) +\mathcal{L}_m \right),
\label{eq:action_1}
\ee
where $V(\phi)$ is the scalar field potential and $\mathcal{L}_m$ represents the Lagrangian of the rest particle spectrum containing both matter and gauge fields. We omit nonminimal couplings of the form $\sim \xi \phi^2 R$, which, although negligible in the early Universe, can become significant at low redshift as the scalar field rolls down, affecting the value of the Planck mass and, consequently, the gravitational constant. Such types of couplings have recently been studied in light of DESI data in Refs.~\cite{Ye:2024ywg,Ye:2024zpk,Wolf:2024stt,Wolf:2025jed,Wolf:2025jlc}.

To model the quintessence dynamics, we adopt a Higgs-like potential given by\footnote{Note that throughout this paper, scalar potential values are expressed in units of $H_0^2\Mpl^2 \simeq (2.28\times10^{-3}\, {\rm eV})^4 h^2$, and the scalar field mass is given in units of $H_0 \simeq 2.13\times 10^{-33}h\, {\rm eV} $, where $h$ is the reduced Hubble parameter.}
\be
    V(\phi) = V_0 + \frac{m^2}{2}  \phi^2 + \frac{\lambda}{4}  \phi^4\,,
\label{eq:higgs_pot}
\end{equation}
which for $m^2<0$, develops a minimum at $\phi=\sqrt{-m^2/\lambda}\equiv v_\phi$. 
The priors used for these parameters in the analysis of the next section are listed in Table~\ref{table:Priors}.

Potentials of this type, albeit with $\lambda=0$, have been previously studied in the context of the DESI data in Refs.~\cite{Wolf:2024eph,Cline:2025sbt} (see also Ref.~\cite{Wolf:2023uno}). Here, we extend this framework by including the renormalizable $\sim\phi^4$ self-interaction term, which stabilises the potential and allows for a richer dynamical behavior of the scalar field. This addition enables a more comprehensive exploration of the late-time evolution of the Universe, including the possibility of transitioning into either a de Sitter (dS) phase, characterized by eternal accelerated expansion (a ``big freeze"), or an anti-de Sitter (AdS) phase, which may ultimately give rise to a cosmological collapse (a ``big crunch")~\cite{Felder:2002jk}. The interplay between the parameters $V_0,\, m^2,$ and $\lambda$ governs the asymptotic fate of the Universe and provides a theoretically motivated framework for investigating the dynamics of dark energy beyond phenomenological parametrizations.

Variations of the action in Eq.~\eqref{eq:action_1} with respect to the scalar field 
and the metric tensor, assuming a spatially flat FRW background, yield the following equations of motion
\begin{subequations}
\begin{align}
    \ddot{\phi} +3H\dot{\phi} 
    & = -\partial_\phi V(\phi)\,, 
    \\
    H^2/H_0^2 
    &= \Omega(z) +(\dot{\phi}^2/2+V(\phi))/ (3H_0^2\Mpl^2)\,,
    \label{eq:fr2}
\end{align}
\end{subequations}
where, as discussed previously, $\Omega(z)$ encapsulates the energy density contributions from matter and radiation, normalized to the critical density. From~\eqref{eq:fr2}, we observe that if the potential $V(\phi)$ becomes negative, it can offset the positive contributions from $\Omega(z)$ and the kinetic energy. When this occurs, the Hubble parameter vanishes, signaling the end of cosmic expansion. Beyond this point, $H$ becomes negative and the Universe enters a phase of contraction. The scalar field will oscillate around the negative minimum of the potential, but will eventually diverge due to the Hubble antifriction, which accelerates its motion. Thus, potentials with a negative minimum have the same qualitative effect on cosmic evolution as those unbounded from below~\cite{Felder:2002jk}.

The effective equation of state parameter of the scalar field is given by
\be
    w(a) = \frac{\dot{\phi}^2/2-V(\phi)}{\dot{\phi}^2/2+V(\phi)}\,.
\ee
In the early Universe, where the field is nearly frozen due to Hubble friction (i.e., $\dot{\phi}\sim0$) and the potential is positive, the equation of state parameter approaches $w\gtrsim-1$, mimicking a cosmological constant. As the Universe evolves and the scalar field begins to roll, the dynamics become increasingly sensitive to the shape of the potential. In particular, since the potential energy can become negative, the equation of state parameter $w(a)$ can evolve across a broad range of values larger than $-1$. Specifically, $w$ increases from its initial value near $-1$, can reach values $w \gg 1$ before the onset of collapse, and eventually approaches $w = +1$, corresponding to a stiff equation of state~\cite{Steinhardt:2001st,Felder:2002jk,Heard:2002dr,Kallosh:2002gf}.

\section{Methodology and results}
\label{sec:meth_res}

We performed a maximal likelihood analysis by adopting the Markov Chain Monte Carlo (MCMC) method.
We use DESI DR2 BAO data from Ref.~\cite{DESI:2025zgx}, which span seven redshift bins covering the range of $z \sim 0.3 - 2.3$. For all but the lowest redshift bin -- which is restricted by a small comoving volume -- DESI provides anisotropic BAO measurements, i.e., along and perpendicular to the line of sight. This results in a total of $2\times 6 + 1 = 13$ BAO data points. For the CMB, we use a radically compressed dataset (e.g., Refs.~\cite{WMAP:2010qai,Bansal:2025ipo,DESI:2025zgx}) consisting of three effective data points, calibrated for Planck observations, as in Ref.~\cite{DESI:2025zgx}. SN data are taken from the DES Year 5 release~\cite{DES:2024jxu}, which contains 1829 distance moduli measurements: 1635 from DES itself, supplemented by 194 nearby SNe Ia ($z<0.1$) from the Harvard-Smithsonian Center for Astrophysics and  Carnegie Supernova Project (CfA/CSP) foundation sample~\cite{Hicken:2009df,Hicken:2012zr,Foley:2017zdq}. Further details of our MCMC analysis are provided in Appendix~\ref{appendixa}.

\begin{figure}[t!]
    \centering
    \includegraphics[width=0.98\linewidth]{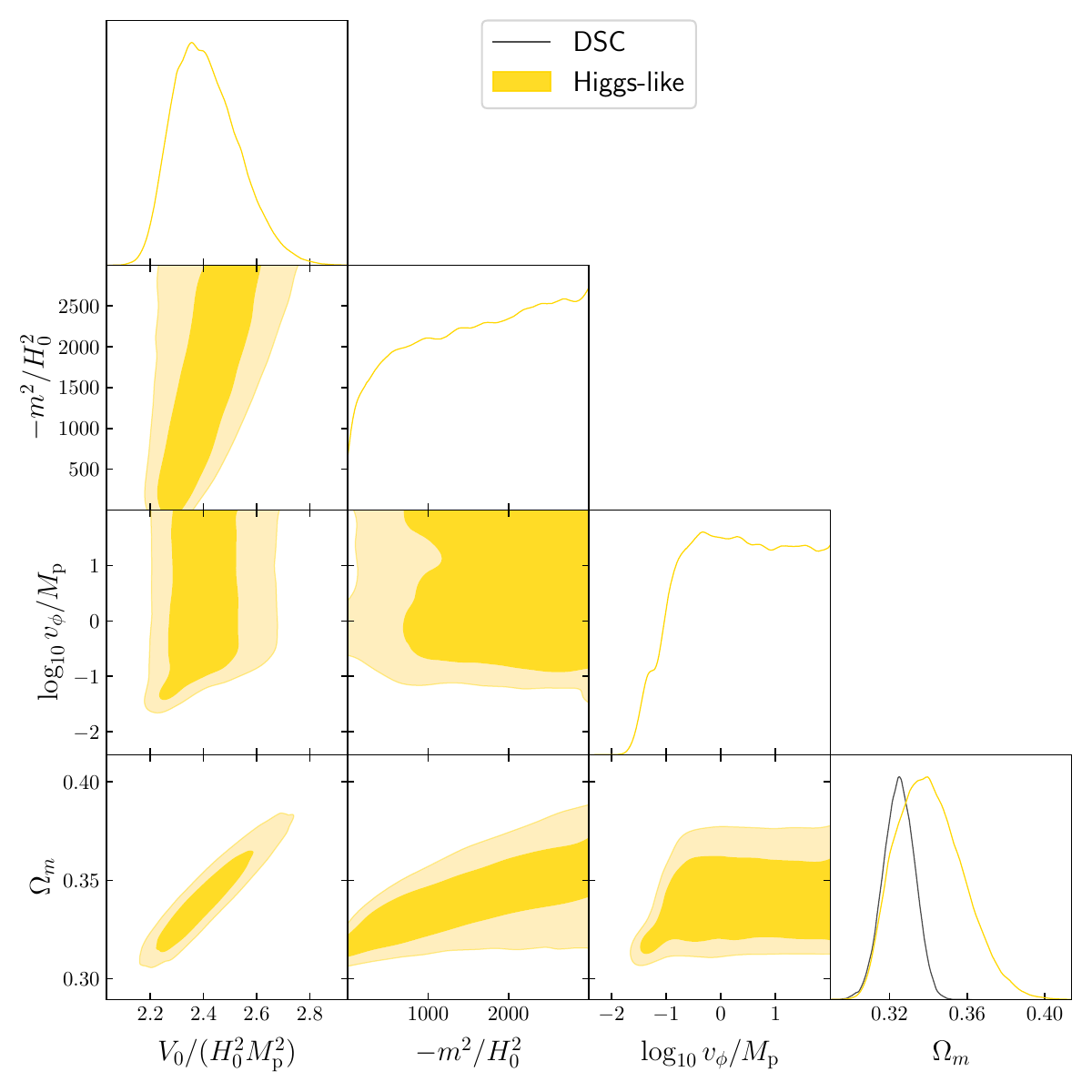}
    \caption{Posterior distributions for the parameters of the Higgs-like potential and effective DSC parametrization. It was assumed that $m^2 < 0$ and $\phi_i<v_\phi$. The light and dark color corresponds to the $68\% $ and $95\%$ confidence level, respectively.} 
    \label{fig:posterior_quint}
\end{figure}

Our main results are presented in Table~\ref{table:CredibleIntervals} and illustrated in Figs.~\ref{fig:evol_z}, \ref{fig:posteriors_CPL} and~\ref{fig:posterior_quint}. In the following, we provide a detailed analysis, beginning with the parametrizations of $w$ and subsequently focusing on the Higgs-like model. 

\subsection{Parametrizations of \texorpdfstring{$w$}{w}}

From Table~\ref{table:CredibleIntervals}, it is clear that the standard $\Lambda$CDM model is disfavored compared to all the considered parametrizations of the equation of state parameter $w$. Specifically, the difference in chi-squared values, $\Delta\chi^2$, increases by $17.2$ for both the CPL and tCPL parametrizations, indicating a similar fit to the data. In terms of the Akaike Information Criterion (AIC)~\cite{1974ITAC...19..716A}, this corresponds to improvements of $13.2$ and $11.2$, respectively. For the tCPL(-1) model, the increase in $\Delta\chi^2$ is $13.5$, with a corresponding $\Delta\mathrm{AIC}$ of $9.5$. The DSC model exhibits a similar level of statistical significance to the tCPL(-1) model, with $\Delta\chi^2 = -13.3$ and $\Delta \mathrm{AIC} =- 9.3$.
\begin{figure}[t!]
    \centering
    \includegraphics[width=1\linewidth]{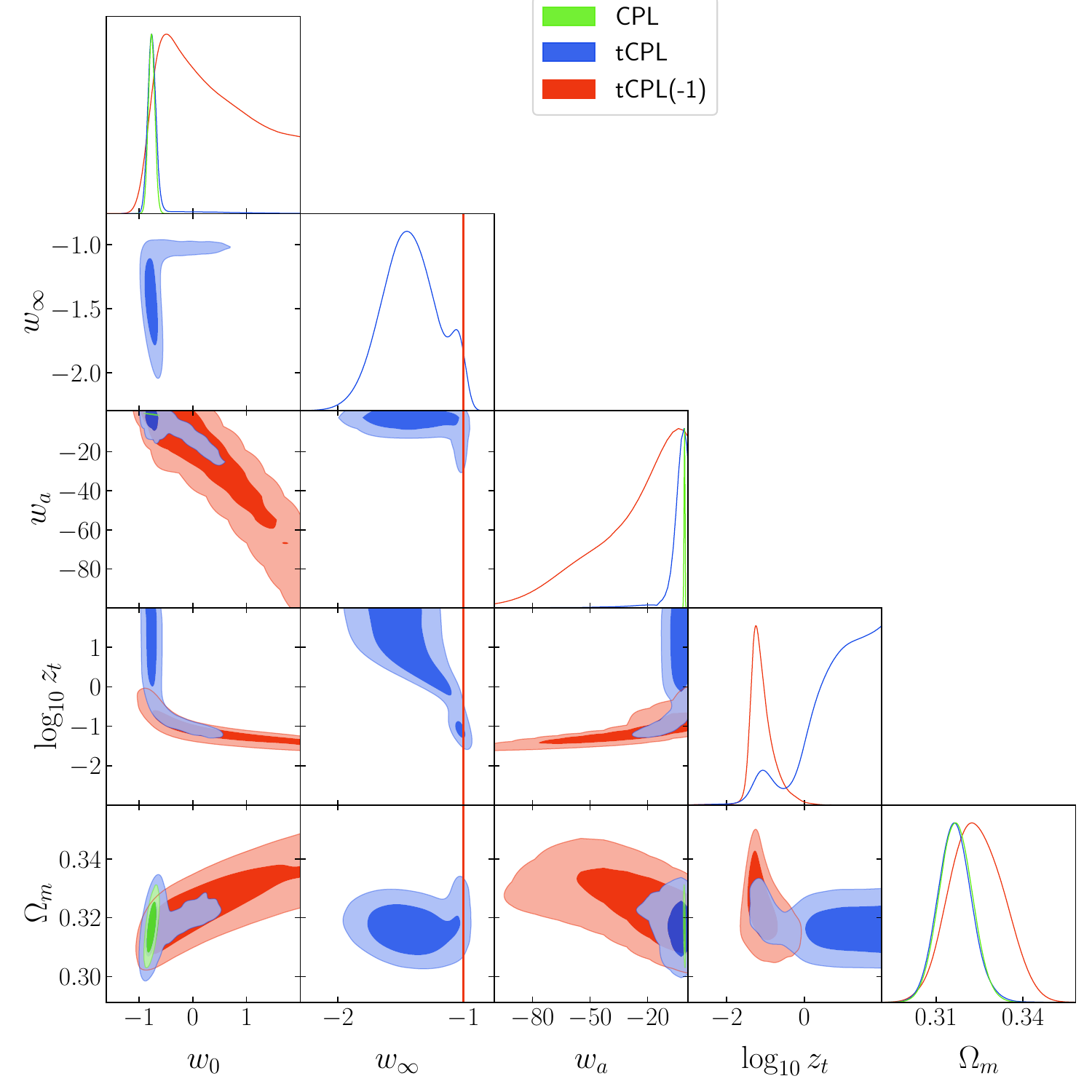}
    \caption{Posterior distributions for the parameters of the CPL, tCPL, and tCPL(-1) models for the fit to the CMB+DESI+DESY5 datasets. The tCPL and tCPL(-1) posteriors are shown for the fit with a uniform prior in $\log_{10} z_t$ (a uniform prior in $w_a$ is shown in Fig.~\ref{fig:posteriors_full_CPL} in Appendix~\ref{appendixa}). Note that only one of the parameters $z_t$ and $w_a$ is free, and the other is derived from the others.}
    \label{fig:posteriors_CPL}
\end{figure}

These results show a consistent trend favouring models with a time-varying equation of state over the constant $w = -1$ of the $\Lambda$CDM scenario.

Among all parametrizations, the tCPL model appears to provide the best fit according to the $\Delta\chi^2$ criterion. However, it is important to note that tCPL includes one additional parameter compared to the other models, which could account for part of its improved performance. 

Regarding the behavior of $w$, the tCPL model predicts an early-time equation of state value $w_\infty$ lying within the ranges $[-1.55, -0.90]$ or $[-1.81, -0.97]$ at the $95\%$ credible intervals, depending on the choice of prior, either on $w_a$ or on $\log_{10}z_t$. The best fit of $w_\infty$ is $-1.45$, with the transition occurring at a redshift of $z_t = 6.45$. These results suggest that phantomlike behavior, where $w < -1$, is generally favored by the data, while the nonphantom case with $w_\infty = -1$ cannot be statistically excluded, as it lies within the $2\sigma$ region (see Fig.~\ref{fig:posteriors_CPL}). Overall, the tCPL parametrization prefers an evolution of $w$ that is qualitatively very similar to the simpler CPL model. Results for the rest of parametrizations are also shown in Table~\ref{table:CredibleIntervals}.

Fig.~\ref{fig:evol_z} illustrates the evolution of the equation of state parameter $w(z)$ (left), the time derivative of the scale factor $\dot{a} = H(z)/(1+z)$ (middle), and the normalized DE density (right). Solid (dashed) lines represent the best-fit (sample mean), while shaded regions indicate the $1\sigma$ credible intervals, clearly showing how the models deviate from the cosmological constant over time.
The only robust and consistent feature across all parametrizations is a sharp drop of $\mathcal{O}(10-15\%)$ in the DE density starting at $z\sim 0.1-1$, which corresponds to a rapid increase in the equation of state parameter.

For quintessencelike models, the DE equation of state begins to deviate significantly from $w=-1$ at redshifts below $z\lesssim 0.1$. This behavior is driven largely by low-redshift SN data. In our previous study~\cite{Gialamas:2024lyw}, we conducted a separate analysis excluding SN data at $z<0.1$, which restored consistency with the $\Lambda$CDM model. In the present work, however, we chose not to pursue this approach further, as $\Lambda$CDM is already in significant tension with CMB and BAO data alone. Moreover, there are independent indications -- in particular, the reported reversal in the sign of the integrated Sachs–Wolfe effect at redshifts $z<0.03$~\cite{Hansen:2025atx} -- suggesting that something genuinely interesting might be occurring at very low redshifts.

From the middle panels, all of the considered models yield a lower $H_0$ compared to $\Lambda$CDM, thereby exacerbating the Hubble tension. This is a common feature among models with a decaying DE density, since they typically lead to higher $\Omega_m$ values, while $\Omega_mh^2$ remains effectively fixed by the CMB. As a result, the Hubble parameter $h$ must decrease. Addressing the Hubble tension in such models requires similar, but more pronounced, modifications as in $\Lambda$CDM, such as altered recombination history, early DE, or other (preferably early-universe) modifications.

Fig.~\ref{fig:posteriors_CPL} shows the posterior distributions for the parameters of the CPL, tCPL, and tCPL(-1) models for the fit to the CMB+DESI+DESY5 datasets. Details of the analysis and the complete corner plots of the posterior distributions are provided in Appendix~\ref{appendixa}, with Fig.~\ref{fig:posteriors_full_CPL} showing the results for the CPL model and its variants, and Fig.~\ref{fig:posteriors_quint_full} presenting those for the DSC parametrization.
\begin{figure}
    \centering
    \includegraphics[width=1\linewidth]{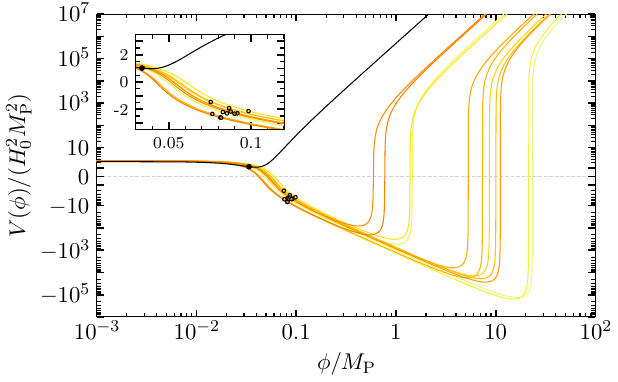}
    \caption{Realizations of the Higgs-like potential for parameters sampled randomly from the $1\sigma$ region of the fit (see Figs.~\ref{fig:posterior_quint},~\ref{fig:posteriors_quint_full}). The solid black line shows the best-fit scenario. The $y$-axis uses an $\rm arcsinh$ scaling, which is linear for small values and logarithmic for large values. The small dots indicate the field values at $z=0$.}
    \label{fig:potential_Higgs}
\end{figure}

What deserves highlighting is that, forbidding phantomlike behavior and imposing $w\geq-1$ as in the tCPL(-1) or DSC parametrizations model, the data will prefer models in which DE begins to evolve at much lower redshifts, that is, $z_t \lesssim 0.1$, as discussed above. Notably, from the posterior of $\log_{10}(z_t)$ in Fig.~\ref{fig:posteriors_full_CPL} one can also observe a degree of bimodality in the tCPL model, as the tCPL (blue) posteriors show a small bump at the same values of $\log_{10}(z_t)$ that are preferred by the truncated tCPL(-1) scenario (red). However, as the tCPL fit shows, without restrictions on phantom behavior, the large $\log_{10}(z_t)$ models are moderately ($\Delta \chi^2 = -3.0$) preferred by the data.

Overall, we observe a degree of bimodality, with the selection favoring two qualitatively distinct classes of models: phantomlike and quintessencelike, as indicated in Table~\ref{table:CredibleIntervals}. Although the phantomlike models provide a slightly better fit, we find that the credible intervals are sensitive to the choice of priors (see Table~\ref{table:CredibleIntervals} and Appendix~\ref{appendixa}). As the preference would ideally be estimated using evidence integrals (rather than just best-fits, as done here), there remains some ambiguity in determining the preferred model. We leave a detailed investigation of the impact of prior selection to future work.

\subsection{Higgs-like quintessence}

In Table~\ref{table:CredibleIntervals}, we present the maximum-likelihood analysis of the CMB+DESI+DESY5 datasets for the Higgs-like potential model defined in Eq.~\eqref{eq:higgs_pot}. The results indicate that the standard $\Lambda$CDM model is disfavored in comparison to the Higgs-like quintessence scenario. Specifically, the corresponding improvement in the fit is quantified by a reduction in the $\chi^2$ value (for $\Lambda$CDM $\chi_{\rm \Lambda CDM}^2 = 1674.29$~\footnote{This $\chi^2$ value is $\sim 3\sigma$ lower than one might naively expect based on the nominal number of degrees of freedom. The low value is entirely driven by the treatment of systematic errors in the SN data, which leads to a significant reduction in the effective number of degrees of freedom, as discussed in~\cite{DES:2024jxu}.}), $\Delta\chi^2 = -13.54$, and a decrease in the Akaike Information Criterion, $\Delta \mathrm{AIC} = -7.5$. Compared to the CPL and tCPL parametrizations, the Higgs-like quintessence model gives a slightly worse fit to the data\footnote{The Higgs-like quintessence model yields a similar $\Delta\chi^2$ to the tCPL(-1) and DSC parametrizations, but results in a larger $\Delta$AIC due to the inclusion of an additional parameter.}, though it performs similarly to the tCPL$(-1)$ case. However, this does not imply that the model is inadequate. On the contrary, it is a simple and well-motivated model inspired by particle physics that can successfully describe the latest observational data. While basic parametrizations such as CPL may match the data slightly better, they are not based on any deeper physical theory and serve only as convenient fitting tools. In contrast, the Higgs-like potential offers a physically consistent and predictive alternative that connects dark energy dynamics to fundamental scalar field theory (see Fig.~\ref{fig:potential_Higgs} for some random realizations of the Higgs-like potential).

In Table~\ref{table:CredibleIntervals}, we also report the best-fit values and the corresponding $95\%$ confidence intervals for the model parameters. For the best-fit point of the MCMC analysis, the scalar field mass is found to be ultralight, $|m|^{\rm (bf)} \simeq 1.16 \times 10^{-31}h$ eV. The vacuum expectation value (VEV) is sub-Planckian, $v_\phi^{\rm (bf)}  \simeq 0.039\,\Mpl$, and the present-day value of the potential is $V^{\rm (bf)} (z=0) = 1.21\, H_0^2 \Mpl^2$. The potential energy at the minimum is $V^{\rm (bf)} (v_\phi) = V_0 + m^2 v_\phi^2 / 4 \simeq 1.15\, H_0^2 \Mpl^2$. Note that in our numerical analysis, we assume a small-field scenario in which the initial field value $\phi_i$ is smaller than the VEV, and the initial velocity is set to zero. Since $H_0$ is fixed, the initial value $\phi_i$ is not a free parameter and must be computed using a shooting method. We find that $\phi_i$ is significantly smaller than the VEV, with a best-fit value of $\log_{10}(\phi_i^{\rm (bf)}/\Mpl) \simeq -46.1$, and a $95\%$ credible interval of $\log_{10}(\phi_i^{\rm (bf)}/\Mpl) \in [-51.5, -13.3]$. Although we do not consider the physics behind the initial value here, this suggests that the initial value may involve some degree of fine-tuning.

As in the case of the parametrizations, Fig.~\ref{fig:evol_z} shows the evolution of $w(z)$, $\dot{a} = H(z)/(1+z)$, and $\rho_{\rm DE}(z)$ for the Higgs-like model, including the best fit, the sample mean and the $1\sigma$ credible interval. Once again, a sharp drop of approximately $\mathcal{O}(13-30\%)$ in $\rho_{\rm DE}(z)$ begins around $z \sim 0.1$. For the best-fit scenario shown in the same figure, we observe that the scalar field has already entered the oscillatory phase, having passed through the minimum of its potential at least once. At present, the value of the equation of state parameter $w$ is approximately $-0.14$.
\begin{figure}[t!]
\centering
\includegraphics[width=1\columnwidth]{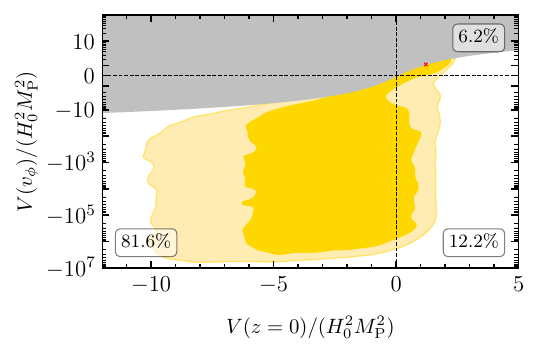}
\caption{Posterior distributions for $V(z=0)$ and $V(v_\phi)$ for the Higgs-like potential~\eqref{eq:higgs_pot}, showing the $68\%$ and $95\%$ confidence regions. The red cross mark indicates the best-fit point. The gray shading shows the mathematically excluded region with $V(z=0) < V(v_\phi)$. The percentages in the corners show the posterior probability of being in the corresponding quadrant. The $y$-axis uses a $\rm arcsinh$ scaling.} 
\label{fig:V_today_V_vaccum}
\end{figure}

Fig.~\ref{fig:posterior_quint} shows the posterior distributions for the parameters $V_0$, $m^2$, and $\log_{10} v_\phi$ of the Higgs-like model, based on the fit to the CMB+DESI+DESY5 datasets. The complete corner plot of the posterior distributions is provided in Fig.~\ref{fig:posteriors_quint_full} in Appendix~\ref{appendixa}. We find that the bounds on the potential parameters are not very restrictive. For the mass and the VEV, we find only the lower bounds $|m^2| > 251 H_0^2$ and $v_\phi \geq 0.08\,\Mpl$. These bounds are almost independent of each other, as we do not observe a strong correlation between $m^2$ and $v_\phi$.

In Fig.~\ref{fig:V_today_V_vaccum}, we illustrate the $68\%$ and $95\%$ confidence intervals for the present value of the potential, $V(z=0)$, versus its value at the minimum, $V(v_\phi)$. Although the best-fit value (red cross mark in Fig.~\ref{fig:V_today_V_vaccum}) suggests that both $V(z=0)$ and $V(v_\phi)$ lie within the dS regime, the data allow for a significant probability that the scalar field may evolve into an AdS phase in the future or that we may already reside in an AdS vacuum today. It is important to note that, regardless of the sign of the potential energy, the total energy density of the Universe today remains positive due to the dominance of the scalar field's kinetic energy.

To quantify these possibilities more precisely, we analyse the explored parameter space and find that the probability of remaining in a dS phase at all times is $6.2\%$, the probability of transitioning from a dS phase today to an AdS phase in the future is $12.2\%$, and the probability of already being in an AdS phase today is $81.6\%$. Both of the latter scenarios imply a future cosmological collapse leading to a reversal of cosmic expansion and an eventual ``big crunch".

\section{Conclusions}
\label{sec:concl}
In this work, we performed a maximum likelihood analysis using the Markov Chain Monte Carlo method to fit data from DESI BAO, the CMB, and supernova observations, to improve the understanding of the evidence for dynamical dark energy. Our primary objective was to investigate not only whether the current observational data remain consistent with the $\Lambda$CDM scenario, which now appears to be increasingly less supported, but also to what extent they can be explained by minimally coupled quintessence or prefer a phantom crossing at high redshifts, that is, $w(z) < -1$.

To explore these possibilities, we fitted the data using extensions of the standard CPL parametrization of the dark energy equation of state. In particular, we considered a truncated version of CPL that behaves as CPL at low redshifts and asymptotically approaches a constant value $w_\infty$ at high redshifts. The best fit for this model yielded $w_\infty \simeq -1.45$, reinforcing the tendency toward phantomlike behavior. However, conclusions regarding the presence of phantom dark energy are not statistically robust, as nonphantom values such as $w_\infty \simeq -1$ are still within the $2\s$ confidence region.

Given the theoretical challenges associated with phantom models, we considered parametrizations that approximate quintessence and explicit, physically motivated models based on scalar fields minimally coupled to gravity. These models can fit the data well without invoking phantom behavior. Specifically, we considered a scalar field with a Higgs-like potential of the form
$ V(\phi) = V_0 + \frac{1}{2} m^2 \phi^2 + \frac{1}{4} \lambda \phi^4,$
which includes the possibility of a nonvanishing vacuum energy, an ultralight negative-squared mass, and a tiny, fine-tuned initial field value $\phi_i$. Our numerical analysis shows that, for the best-fit parameters, the vacuum energy is positive and the scalar field has already begun to oscillate around its minimum. Nevertheless, we found that a negative vacuum energy is preferred at the $\sim 93.8\%$ level. If this scenario holds, a future cosmological collapse leading to a reversal of the cosmic expansion would be unavoidable. It is important to note that the Hubble tension cannot be addressed in the context of our models, as they all yield $w(z=0) > -1$, which inevitably leads to an $H_0$ lower than the corresponding $\Lambda$CDM value.

Finally, our results indicate that the sharp drop in dark energy density is a recent effect appearing around $z \sim 0.1$. However, because the observable Universe at these redshifts is limited by causality, it remains unclear whether this is a global property of dark energy or a local fluctuation. One possible explanation, as also discussed in our previous study~\cite{Gialamas:2024lyw}, is that this feature could be driven by the low-$z$ supernova data. Today’s supernova samples are large enough that statistical errors are often surpassed by systematic uncertainties, which are difficult to fully control due to the complex and not yet fully understood physics of supernovae. Moreover, for nearby SN, the analysis is further complicated by the peculiar motions of their host galaxies, which can significantly bias distance estimates if not accurately modelled.

\begin{acknowledgments}
\vspace{5pt}\noindent\emph{Acknowledgments --}
This work was supported by the Estonian Research Council grants PSG869, MOB3JD1202, RVTT3 and RVTT7 and the Center of Excellence program TK202 ``Fundamental Universe''. This article is based upon work from COST Action CA21136 – “Addressing observational tensions in cosmology with systematics and fundamental physics (CosmoVerse)”, supported by COST (European Cooperation in Science and Technology).
\end{acknowledgments}

\vspace{5pt}\noindent\emph{Note added -- }
While this article was being prepared for submission, the work~\cite{Ozulker:2025ehg} appeared, presenting a study of the tCPL(-1) parametrization and discussing a possible crossing into the phantom regime. That study excludes the absence of a phantom crossing at a significance above $3\sigma$.

\appendix
\section{Details on the numerical analysis}
\label{appendixa}

To sample the posteriors for the different models considered in this study, we have adopted the MCMC method. The MCMC analysis was performed using the Python package \textbf{emceee}~\cite{Foreman-Mackey:2012any}, which uses an affine-invariant ensemble sampler. The priors for model parameters are listed in Table~\ref{table:Priors}. The resulting posteriors are displayed with the \textbf{getdist} package~\cite{Lewis:2019xzd}. Further analysis was conducted using \textbf{arviz}~\cite{Kumar2019}, and to determine the best-fit points and the $\chi^2$ value, we employed \textbf{iminuit}~\cite{iminuit}. Chain convergence was checked and two criteria were used: Gelman-Rubin statistics $\hat{R} < 0.01$ or effective sample size to be $ \rm{ESS}> 1000$. 

For tCPL sampling and tCPL(-1) using $w_a$ as sampling parameter, the convergence was not reached. For tCPL (flat $\log_{10}z_t$) the worst effective sample size was $\approx 600$ and for tCPL (flat $w_a$) was $\approx 110$. This is caused by multimodality. For tCPL(-1) (flat $w_a$) the ESS was $\approx 360$.

\begin{figure*}[ht!]
    \centering
    \includegraphics[width=1\linewidth]{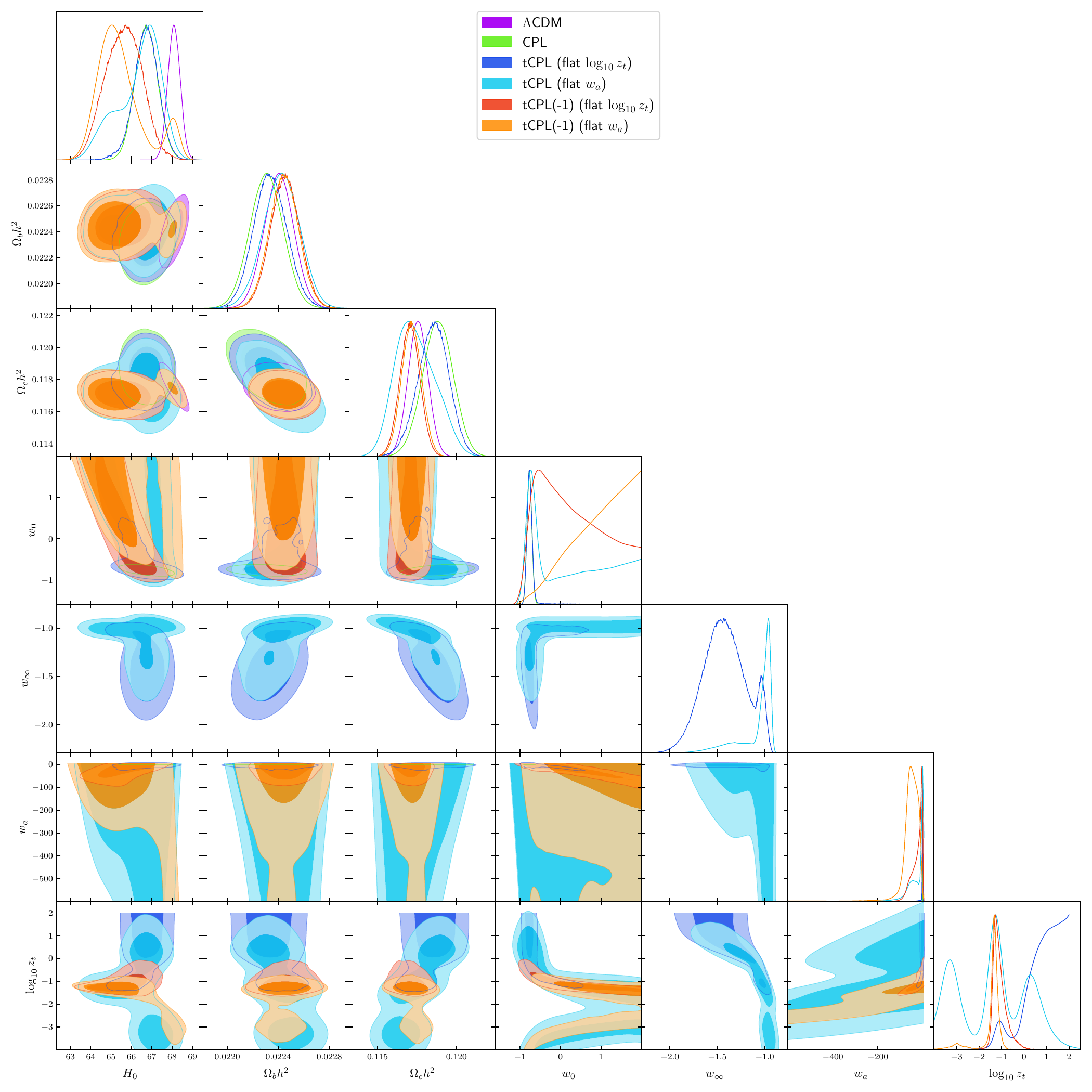}
    \caption{Posterior distributions for the parameters of the CPL, tCPL, and tCPL(-1) models for the fit to the CMB+DESI+DESY5 datasets. The tCPL and tCPL(-1) posteriors are shown for the fit with a uniform prior in $\log_{10} z_t$ (blue and red) as in Fig.~\ref{fig:posteriors_CPL}, and a uniform prior in $w_a$ (light blue and orange). Note that only one of the parameters $z_t$ and $w_a$ is varied, and the other is derived.}
    \label{fig:posteriors_full_CPL}
\end{figure*}

\begin{figure*}[t!]
    \centering
    \includegraphics[width=\linewidth]{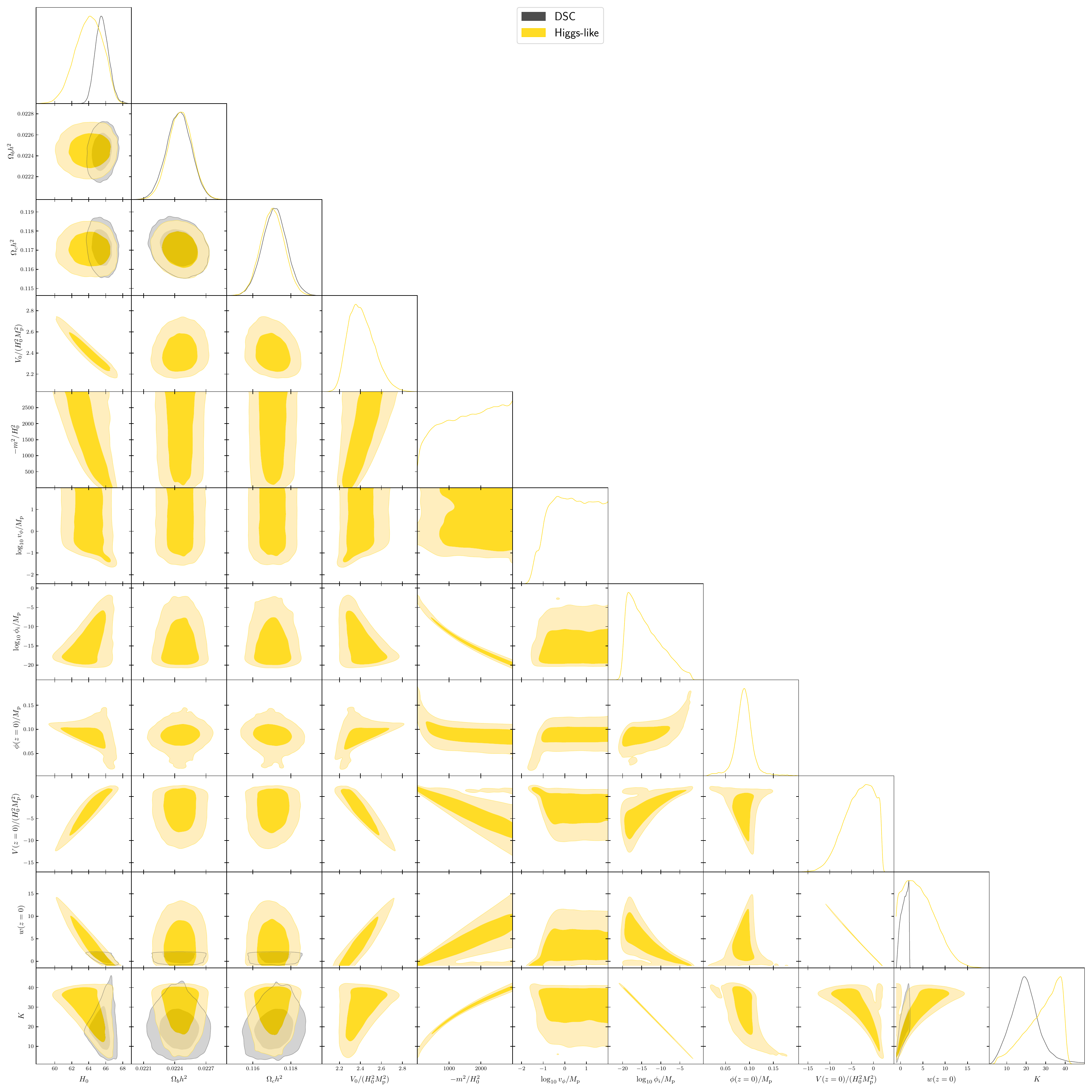}
    \caption{Posterior distributions for the parameters of the Higgs-like potential and effective DSC parametrization. It was assumed that $m^2 < 0$ and $\phi_i<v_\phi$.} 
    \label{fig:posteriors_quint_full}
\end{figure*}

For completeness, the posteriors for all varied model parameters, including some derived parameters, are listed in Figs.~\ref{fig:posteriors_full_CPL}, and~\ref{fig:posteriors_quint_full}.

Following Ref.~\cite{DESI:2025zgx}, the CMB fits were obtained using compressed data, with only three parameters $\theta_*$, $\Omega_{\rm b} h^2$ and $\Omega_m h^2  = (\Omega_{\rm b} + \Omega_{\rm c})h^2$, compared to the model predictions using Gaussian likelihoods. This method allows to pass computationally heavy linear perturbation evolution calculation and works sufficiently well compared to full CMB  analysis~\cite{DESI:2025zgx}.  Fits for DESI DR2 BAO measurements are calculated by using data from~\url{https://github.com/CobayaSampler/sn_data} which provides values for six $(D_M/r_d, D_H/r_d)$ pairs and necessary covariance matrices. For one tracer, we have access only to $D_V/r_d$ (and its covariance matrix). These measurement values, in combination with model predictions, are used to calculate Gaussian likelihoods. The detailed description of calculating fits for DES-Y5 data is given in Ref.~\cite{Goliath:2001af}. The latter method requires the calculation of the model prediction for the luminosity distance $D_{L} = (1+z)D_M(z)$, which is used in combination with data points from~\url{https://github.com/des-science/DES-SN5YR} to calculate likelihoods for DES-Y5 results.

When modeling cosmological expansion, we include the contribution of neutrinos using the approach developed in Ref.~\cite{WMAP:2010qai} and further elaborated upon in Ref.~\cite{Bansal:2025ipo}. Assumed two massless and one massive neutrino ($m_\nu = 0.06 \rm eV$), we have
\bea
    \Omega_\nu(a) 
    \!=\! 0.2271 \,\Omega_\gamma (a) N_{\rm{eff}} \left(\frac{2 + f(m_\nu a /T_{\nu, 0})}{3}\right)\,
\eea
and we use the approximation $f = \left(1+(Ay)^p\right)^{1/p}$,  with $A = 180 \zeta(3)/(7\pi^4) \simeq 0.3173$ and $p = 1.83$~\cite{WMAP:2010qai}.

The sound horizon at redshift $z$ is
\be
    r(z) = \int \limits_{z}^{\infty} \td z'\, \frac{c_s(z')}{H(z')}\,.
\ee 
For BAO, the redshift $z=z_d$ is fixed by the end of the drag epoch.  This integral can be approximated and calibrated~\cite{Brieden:2022heh} using Planck satellite measurements~\cite{Planck:2018vyg}, as follows
\be
    r_d \!=\! 147.05 {\rm Mpc} \!
    \left[\frac{\Omega_b h^2}{0.02236}\right]^{-0.13}\! 
    \left[\frac{\Omega_{m} h^2}{0.1432}\right]^{-0.23}\!
    \left[\frac{N_{\rm{eff}}}{3.04}\right]^{-0.1}
    \!\!\!\!\!\!\!\!.
\ee
For CMB, we use the angular scale $\theta_* = D_M/r_{*}$. Here $r_*$ is the sound horizon fixed at the end of the recombination $z_* \approx 1089$. Further, $z_*$ depends on the underlying cosmology, but as we investigate only late dark energy, the change should be minimal.

For BAO data comparison, we have a pair of values: transverse comoving distances 
\be
    D_M(z) = \frac{1}{H_0} \int\limits_{0}^{z} \frac{\td z'}{E(z')}\,,
\ee
and Hubble length $D_H(z) = 1/H(z)$. For a single BAO datapoint, only the isotropic BAO distance $D_V \equiv (zD_M(z)^2D_H(z))^{1/3}$ is used.

\onecolumngrid

\setlength{\tabcolsep}{10.pt}
\renewcommand{\arraystretch}{1.8}
\begin{table}[t!]
  \centering
  \begin{tabular}{|c|c|cc|ccc|}
    \hline 
    \hline 
  & & \multicolumn{2}{c|}{\textbf{phantomlike}} & \multicolumn{3}{c|}{\textbf{Quintessencelike}} 
  \\
    \hline \\[-0.67cm]    
    Variable & \bf{$\Lambda$CDM} & \bf{CPL} & \bf{tCPL}   &  \bf{tCPL(-1) } & \bf{DSC} & \bf{Higgs-like}\\
    \hline\\[-0.67cm] 
    $H_0$               & $[40, 90]$      & $[40, 90]$      & $[40, 90]$           & $[40, 90]$ & {$[40, 90]$} & $[50, 90]$ \\ 
    $\Omega_b h^2$      & $[0.005, 0.1]$ & $[0.005, 0.1]$ & $[0.005, 0.1]$ &  $[0.005, 0.1]$ & {$[0.005, 0.1]$} & $[0.005, 0.1]$\\ 
    $\Omega_c h^2$      & $[0.002, 0.2]$  & $[0.002, 0.2]$  & $[0.002, 0.2]$  &  $[0.002, 0.2]$ & {$[0.002, 0.2]$} & $[0.002, 0.2]$ \\ 
    $w_0$               & ---    & $[-3, 2] $    & $[-3, 2]$ &  $[-3, 2]$ & {$[-1, 2]$} & --- \\ 
    $w_a$      & ---    & $ [-6006, 2]$    & $$[-6006, 5]$$ &  $[-3003, 2]$ & --- & --- \\ 
    $\log_{10}z_t$    & ---    & ---    & $[-3, 2]$ &  $[-3, 2]$ & ---& --- \\ 
    $w_\infty$    & ---    & ---    & $[-4, 2]$ &  --- & ---& --- \\ 
    $K$ & ---    & ---    & --- &  --- & {$[1, 80]$}& --- \\
    $V_0/\Mpl^2H_0^2$ & ---    & ---    & --- &  --- & --- & $[0, 6]$\\
    $-m^2/H_0^2$ & ---    & ---    & --- &  --- & --- & $[0, 3000]$\\
    $\log_{10}( v_\phi/\Mpl)$ & ---    & ---    & --- &  --- & --- & $[-5, 2]$\\
    \hline
    \hline
  \end{tabular}
  \caption{Priors for the parametrizations and the Higgs-like model introduced in the main text. For the tCPL and tCPL(-1) models, the prior $w_a$ or $\log_{10} z_t$  is used. For the truncated models, we additionally impose the condition $w_0 + w_a < w_\infty < w_0$.}
  \label{table:Priors}
\end{table}

\bibliographystyle{JHEP}
\bibliography{dark}
\end{document}